\documentclass[pra,twocolumn,epsfig,rotate,superscriptaddress,showpacs]{revtex4}
\usepackage{graphicx}
\usepackage{epstopdf}
\usepackage{amsfonts}
\usepackage{braket}
\usepackage{subfigure}
\usepackage{prettyref}
\usepackage{float}
\usepackage{color,CJK}
\usepackage{amssymb}
\usepackage{esint}

\usepackage{braket}

\def\be{\begin{equation}}
\def\ee{\end{equation}}
\def\ba{\begin{eqnarray}}
\def\ea{\end{eqnarray}}

\begin{document}

\title{Monopoles  in non-Hermitian systems}
\begin{CJK}{UTF8}{gbsn}
\author{Qi Zhang(张起)}
\affiliation{College of Science, Zhejiang University of Technology,
Hangzhou 310023, China}
\author{Biao Wu(吴飙)}
\affiliation{International Center for Quantum Materials, Peking University, 100871, Beijing, China}
\affiliation{Collaborative Innovation Center of Quantum Matter, Beijing 100871, China}
\affiliation{Wilczek Quantum Center, School of Physics and Astronomy, Shanghai Jiao Tong University, Shanghai 200240, China}
\date{\today}
\begin{abstract}
The monopole for the geometric curvature is studied for non-Hermitian systems. We find that the monopole contains not only
the exceptional points but also branch cuts. As  the mathematical choice of branch cut in the complex plane is rather arbitrary,
the monopole changes with the branch-cut choice. Despite this branch-cut dependence, our monopole is invariant
under the $GL(l,\mathbb{C})$ gauge transformation that is inherent in non-Hermitian systems.
Although our results are generic, they are presented in the context of a two-mode non-Hermitian Dirac model.
A corresponding two-mode Hermitian system is also discussed to illustrate the essential difference
between monopoles in Hermitian systems and non-Hermitian systems.
\end{abstract}

\maketitle
\end{CJK}

\section{Introduction}
The monopole is defined as the source of a vector field. In physics,  it was first discussed in electrodynamics.
The electric monopole such as electron and proton exists everywhere in nature while the magnetic one dual to the electric monopole
was first suggested and calculated theoretically by Dirac~\cite{Diracmono}. The field emanating from monopole
becomes divergent or discontinuous at the monopole but is continuous and described by the field flux off the monopole.
The charge of a monopole is well-defined and can be obtained by integrating the field off the monopole.

The magnetic monopole is yet to be discovered. Researchers have instead explored monopoles in the context of geometric phase in quantum system,
where the Berry curvature resembles the magnetic field and the energy degenerate point resembles the monopole~\cite{Berry1,Niu}.
The Berry curvature can exert a Lorentz-like force (often called geometric force) on the electric-neutral particle~\cite{Niu,Mead,hybrid}.
Despite their similarity, Berry curvature has a key difference from the real electromagnetic field:
Berry curvature is  in fact multi-valued in the parameter space with each value associated with a particular energy eigenstate of the Hamiltonian.
In Hermitian systems, the eigenstates with different eigenenergies are well separated from each other. The monopoles of Berry curvature
are defined as the degenerate points, which are the only places where
different bands of eigenstates  can switch smoothly into each other.  In Hermitian systems, once the Hamiltonian is specified,
its monopoles are uniquely determined.

There have been tremendous interests recently in the non-Hermitian systems both theoretically ~\cite{Bender,Bender2,Bender3,Mostafazadeh,Berry,Longhi,West,Mostafazadeh2,Bender4,Bender5,Mostafazadeh3,Mostafazadeh4,Wang,chen,ZhangNJP} and experimentally~\cite{e1,e2,e3,e4,e5,e6,e7,e8,e9,e10,e11,e12,e13,J1,J3,J4,J5}. In non-Hermitian systems, it is known that one band of eigenstates
can smoothly switch to another band without crossing any degeneracy
points (or exceptional points as called in non-Hermitian systems) along a closed loop in the parameter space~\cite{Moiseyev}.
Mathematically, this is a M\"obius loop. In this work we find that, as the result of  this kind of  the ``M\"obius" structure,  the monopoles
in non-Hermitian systems not only contain exceptional points but also branch cuts.
As is well known, branch cuts in the complex plane are not unique mathematically and can be chosen rather arbitrarily.
As a result, the monopoles for a given non-Hermitian system are not uniquely determined by the Hamiltonian. This is in stark contrast
with Hermitian systems. In this work, for simplicity, our results are presented with a two-mode non-Hermitian Dirac model,
where the Chern number, the total charge of the monopole, and its relation to the choice of branch cuts is also discussed.
For comparison,  a two-mode Hermitian system is also studied.

\section{M\"obius loops of Hermitian systems}
\begin{figure}[t]
\includegraphics[width=0.95\linewidth]{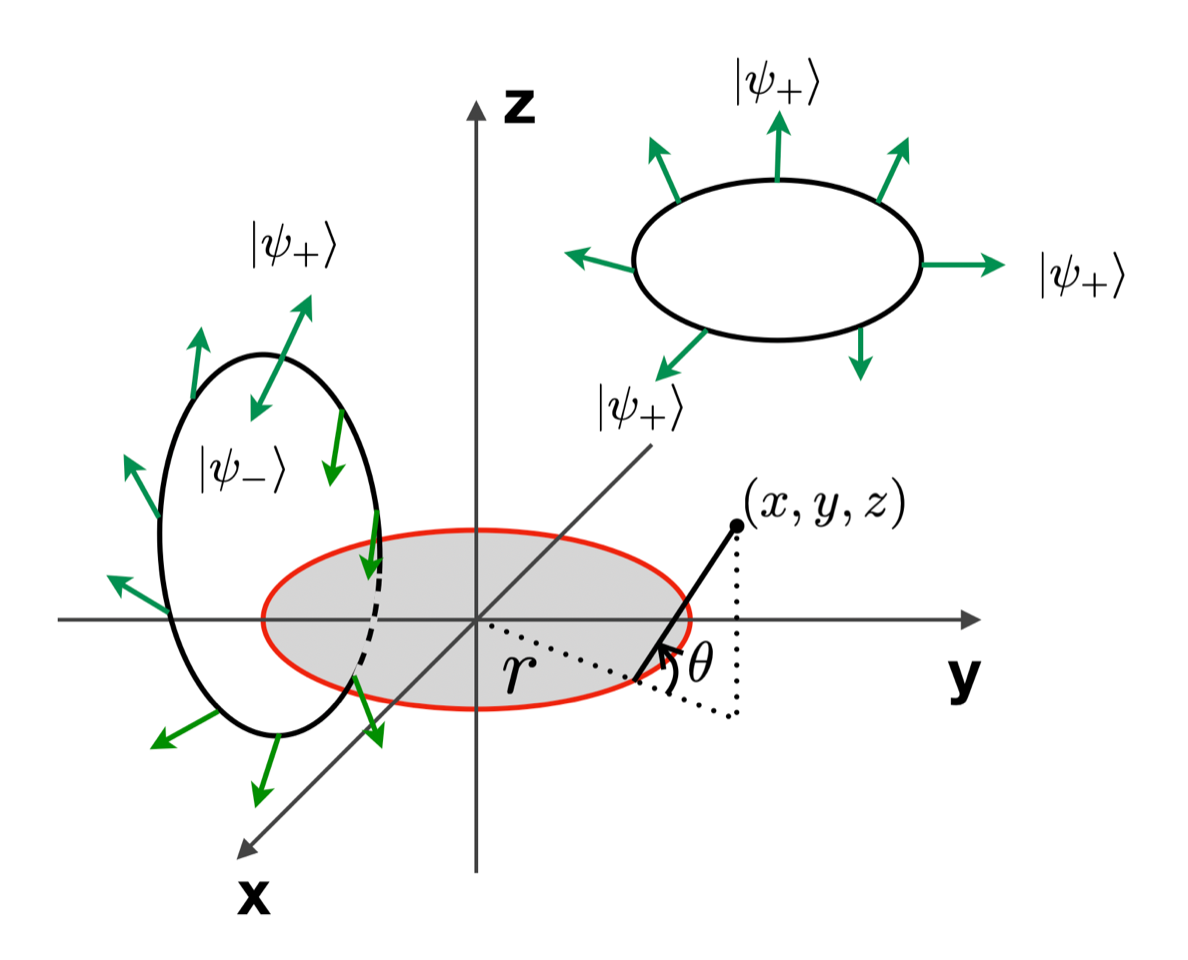}
\vspace*{-0.5cm}
\caption{(color online) Illustration of the M\"obius  loop  in the parameter space of a Hermitian system.
The disk (shaded area with red circle) is the monopole where the system is degenerate.
Two typical loops are shown.  The left loop passes  through the monopole disk  and is a M\"obius loop, along which
the eigenstate  $\ket{\psi_+}$ transits continuously to the other eigenstate $\ket{\psi_-}$ upon return.
 The top right  is an ordinary loop, along which  the eigenstate  $\ket{\psi_+}$ stays the same and
 does not change to $\ket{\psi_-}$. The arrows are drawn schematically to show the directions of eigenstates
 in the Hilbert space.  }
\label{disk}
\end{figure}

We consider a two-mode system, which is described by the following Hamiltonian
\begin{equation} \nonumber \label{para}
H_e(\theta)=\cos\left(\frac{\theta}{2}\right)\left[\cos\left(\frac{\theta}{2}\right)\sigma_z+\sin\left(\frac{\theta}{2}\right)\sigma_x\right],
\end{equation}
where $0\leq\theta\leq2\pi$ is related to a point in a three-parameter space as (see Fig. \ref{disk})
\begin{equation}
\label{sita}
\theta=\left\{\begin{array}{lcl}\arccos\left(\frac{\sqrt{x^2+y^2}-r}{\sqrt{(\sqrt{x^2+y^2}-r)^2+z^2}}\right) & {\rm if}& z\geq0 \\
2\pi-\arccos\left(\frac{\sqrt{x^2+y^2}-r}{\sqrt{(\sqrt{x^2+y^2}-r)^2+z^2}}\right) & {\rm if}& z<0 \end{array} \right.
\end{equation}
It is clear that $\theta=\pi$  is for all the points on the shaded disk in Fig. \ref{disk}, which is mathematically
specified by  $x^2+y^2\le r^2,z=0$.

The eigen-energies of $H_e$ is $E_{\pm}(\theta)=\pm\cos\left(\frac{\theta}{2}\right)$ and the corresponding eigenstates are
\begin{equation}
\ket{\psi_+(\theta)}=\pmatrix{\cos\frac{\theta}{4}\cr \sin\frac{\theta}{4}}\,,
~~\ket{\psi_-(\theta)}=\pmatrix{-\sin\frac{\theta}{4}\cr \cos\frac{\theta}{4}}
\end{equation}
As $E_{\pm}(\pi)=0$, the system is degenerate on the whole shaded disk (see Fig. \ref{disk}).
It is clear that the two energy bands $E_{+}(\theta)$ and $E_{-}(\theta)$ can switch to each other
smoothly only at these degenerate points. In other words, the disk is the monopole of
$H_e$;  this is in stark contrast with the usual case, where the monopole is  a single point.
Moreover, we notice that
\begin{equation}
\ket{\psi_+(\theta+2\pi)}=\ket{\psi_-(\theta)}\,.
\end{equation}
This interesting fact implies the existence of M\"obius loops in our systems. Consider the  lower-left loop in Fig. \ref{disk}.
It is clear that if we start with a point corresponding to an angle $\theta$ and  traverse along the loop, we will end up with angle
$\theta\pm 2\pi$ when we are back at the same point. This means that along this loop  if we start with eigenstate
$\ket{\psi_+}(\theta)$ at a given point, we will end up with  $\ket{\psi_-}(\theta)$  when we are back at the same point.
This is    a M\"obius loop. In fact, any loop that intersects with the monopole disk is a M\"obius  loop.
Loops that do not intersect with the monopole disk are not  M\"obius  loops.
 Note  that in general we can set an arbitrary point along a M\"obius loop where the two set of vectors on the loop switch to each other.
 Here it is convenient and natural to choose the degenerate points.

The corresponding physics  is clear. Along a M\"obius  loop there is an energy degenerate point, where one can smoothly
go from one energy $E_+(\theta)$ to the other $E_-(\theta)$ or vice versa. Along an ordinary loop, there is no energy degenerate point
and one can only stay in one eigen-energy if the parameter changes slowly along the loop. In the next section, we will see that
it is possible to have exact M\"obius loops that do not contain energy degenerate points in non-Hermitian systems.

\section{M\"obius loop in non-Hermitian systems}

For  a non-Hermitian Hamiltonian $H$, the eigen-energies are  in general complex. There are usually  exceptional
points (EPs) in the parameter space at which eigen-energies  are degenerate.  At EPs,
the  non-Hermitian Hamiltonian $H$ is non-diagonalizable; off EPs,  $H$ is diagonalizable and
admits a set of biorthonormal eigenvectors $|\psi_j\rangle$ and $|\phi_j\rangle$~\cite{F3}, satisfying,
\begin{eqnarray} \label{eigen}
& H|\psi_n\rangle=E_n|\psi_n\rangle, ~~~&
H^\dag|\phi^n\rangle=E_n^*|\phi^n\rangle, \\  \label{right-left}
& \langle\phi^m|\psi_n\rangle=\delta_{mn}, ~~~&
\sum_n |\psi_n\rangle\langle\phi^n|=1.
\end{eqnarray}
The eigenstates $|\psi_j\rangle$ and $|\phi^j\rangle$ can be regarded as  contravariant and covariant vectors in Hilbert space,
respectively~\cite{ZhangWu}.

For simplicity and without loss of generality we consider a simple non-Hermitian Hamiltonian~\cite{T1,T2,T3,T4,T5,T6,T7,T8,T9,Fu,chen1,chen2},
\begin{equation} \label{gene-H}
H=p_x\sigma_x+p_y\sigma_y+(p_z+\text{i}s)\sigma_z\,.
\end{equation}
The  parameter space here is spanned by $\mathbf{p}=(p_x,p_y,p_z)$.  The real parameter $s$ controls
the non-Hermiticity of the system.
When applied to lattice systems, $p_x$, $p_y$ and $p_z$  can be regarded as the Bloch momenta~\cite{Fu,chen1}.

\begin{figure}[t]
\includegraphics[width=0.95\linewidth]{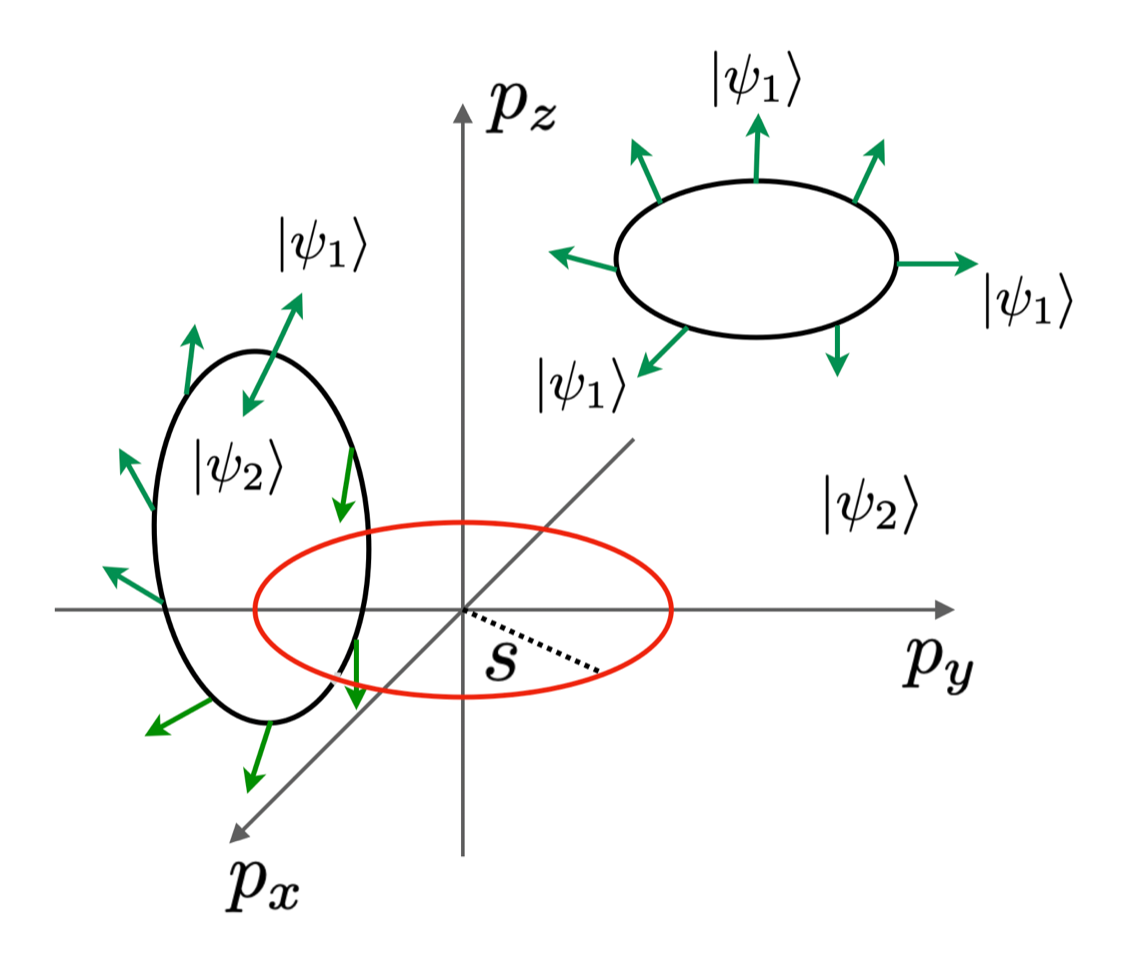}
\vspace*{-0.5cm}
\caption{(color online) Illustration of the M\"obius loop  in the parameter space of a non-Hermitian system.
The red circle is the exceptional points,  where the system is degenerate.
Two typical loops are shown.  The left loop is linked with the red circle  and is a M\"obius loop, along which
the eigenstate  $\ket{\psi_1}$ transits continuously to the other eigenstate $\ket{\psi_2}$ upon return.
 The top right  is an ordinary loop, along which  the eigenstate  $\ket{\psi_1}$ stays the same and
 does not change to $\ket{\psi_2}$. The arrows are drawn schematically to show the directions of eigenstates
 in the Hilbert space.  }
\label{circle}
\end{figure}

The eigen-energies  of $H$  can be easily obtained
\begin{equation} \label{Gen}
E_{1,2}=\pm\sqrt{\mathbf{p}^2-s^2+2\text{i}p_zs}\,.
\end{equation}
It is clear that $E_1=E_2=0$ at the circle of radius $s$ that is given by $p_x^2+p_y^2=s^2,~p_z=0$,
and $E_1\neq E_2$ at other points in the parameter space. So, the EPs form a  circle.
The corresponding biorthonormal eigenstates  can be worked out,
respectively, as (up to a $GL(l,\mathbb{C})$ gauge)~\cite{ZhangWu},
\begin{eqnarray}  \label{wf1}
|\psi_1\rangle&=&\left(\begin{array}{c} \sqrt{\mathbf{p}^2-s^2+2\text{i}p_zs}+is+p_z \\ \\ p_x+\text{i}p_y  \end{array}\right) \\
|\psi_2\rangle&=&\left(\begin{array}{c}  -\sqrt{\mathbf{p}^2-s^2+2\text{i}p_zs}+is+p_z  \\ \\p_x+\text{i}p_y \end{array}\right)
\end{eqnarray}
\begin{eqnarray} \label{wf2}
|\phi^1\rangle&=&\left(\begin{array}{c} \frac{1}{2\sqrt{\mathbf{p}^2-s^2-2\text{i}p_zs}} \\ \\
\frac{\sqrt{\mathbf{p}^2-s^2-2\text{i}p_zs}+\text{i}s-p_z}{2(p_x-\text{i}p_y)\sqrt{\mathbf{p}^2-s^2-2\text{i}p_zs}} \end{array}\right)
\\
|\phi^2\rangle&=&\left(\begin{array}{c}  -\frac{1}{2(\sqrt{\mathbf{p}^2-s^2-2\text{i}p_zs}
-\text{i}s+p_z)}\\ \\ \frac{1}{2(p_x-\text{i}p_y)}
 \end{array}\right)
\end{eqnarray}

As the eigen-energies $E_{1,2}$ are the square roots of a complex variable, there is a branch cut:
when the parameters $p_x$, $p_y$, and $p_z$ change continuously along a loop that goes through the branch cut,
$E_1$ changes to $E_2$ and $E_2$ changes to $E_1$. This is naturally a M\"obius loop.
The left loop as shown in Fig. \ref{circle} is such a M\"obius loop. In fact, all the loops linked with the red circle,
where the EPs are located, are M\"obius loops.  All the loops that are off
the red circle are not M\"obius loop; one example is shown in the top right corner of Fig. \ref{circle}.
As the function $E_{1,2}=\pm\sqrt{\mathbf{p}^2-s^2+2\text{i}p_zs}$ also appears in the corresponding eigenstates,
the eigenstates will also switch to each other at the branch cut along a M\"obius loop.

However, there are crucial differences between Hermitian systems and non-Hermitian systems for M\"obius loops.
In the Hermitian case, the M\"obius loop has to go through a degenerate point; in the non-Hermitian case,
the M\"obius loop  only needs to link with the circle of exceptional points and there is no exceptional point
on the M\"obius loop.  Moreover, in the Hermitian case, the two eigen-energies $E_{\pm}$ switch to
each other at degenerate points, which are fixed once the Hamiltonian is given;
in the non-Hermitian case, the two eigen-energies $E_{\pm}$ switch to each other
at the branch cut, which is to be chosen arbitrarily even the Hamiltonian is given. The disk $p_x^2+p_y^2\leq s^2,~p_z=0$ is a
natural choice of branch cut; however, as a well-known mathematical fact,
we can choose other branch cuts. As a result, in the Hermitian system, the monopole consists of only degenerate points;
in the non-Hermitian system, the monopole consists of  exceptional  points and branch cuts as we explain in detail next.


\section{Monopoles in non-Hermitian Hamiltonians}
In  non-Hermitian systems,
the Berry connection and Berry curvature for the $j$th eigenstate are given by~\cite{ZhangWu,GarrisonPLA},
\begin{eqnarray} \label{Gcon}
\mathbf{A}_j&=&\text{i}\langle\phi^j|\nabla|\psi_j\rangle;  \\ \label{Gcur}
\mathbf{B}_j&=&\text{i}\langle\nabla\phi^j|\times|\nabla\psi_j\rangle,
\end{eqnarray}
where $\nabla\equiv\frac{\partial}{\partial\mathbf{R}}$ with $\mathbf{R}$ being the adiabatic parameters. The Berry connection $\mathbf{A}_j$ for a non-Hermitian eigenstate is up to a $GL(l,\mathbb{C})$ gauge $\mathbf{A}_j'=\mathbf{A}_j+\text{i}\frac{1}{f}\nabla f$ upon the $GL(l,\mathbb{C})$ gauge transformation of the biorthonormal eigenstate~\cite{ZhangWu},
\begin{equation} \label{gauge55}
|\psi_j'\rangle=f|\psi_j\rangle, \; \langle\phi^{'j}|=\frac{1}{f}\langle\phi^j|,
\end{equation}
with $|f|\neq1$ and $f\in GL(1,\mathbb{C})$. However, as in the case of Hermitian eigenstate, the Berry curvature is invariant upon the gauge transformation.

In our case of Hamiltonian (\ref{gene-H}), $\mathbf{R}$ are $p_x,p_y,p_z$ and
the Berry curvature can be worked out as,
\begin{eqnarray}  \label{Berry-Cur}
\mathbf{B}_1&=&\text{i}\langle \nabla\phi^1|\times|\nabla\psi_1\rangle \\   \nonumber &=&-\frac{\mathbf{p}}{2(\mathbf{p}^2-s^2+2\text{i}p_zs)^{3/2}}-\frac{\text{i}s}{2(\mathbf{p}^2-s^2+2\text{i}p_zs)^{3/2}}\hat{p}_z, \\ \nonumber
\mathbf{B}_2&=&\text{i}\langle \nabla\phi^2|\times|\nabla\psi_2\rangle=-\mathbf{B}_2.
\end{eqnarray}
where $\mathbf{p}\equiv(p_x,p_y,p_z)$, $\hat{p}_z$ is the unit vector in $p_z$ direction and $\nabla\equiv\frac{\partial}{\partial\mathbf{p}}$.
At $s=0$, the result reduces apparently to that for the standard spinor.

Before focusing on Hamiltonian (\ref{gene-H}), we present a general discussion on Berry curvature $\mathbf{B}_j$.
Let us calculate the divergence of the Berry curvature, i.e. $\nabla\cdot\mathbf{B}_j$.
We introduce an auxiliary operator
\begin{equation} \label{e37}
\mathbf{F}=-{\text i}\sum_{n}|\nabla \psi_{n}\rangle\langle \phi^{n}|={\text i}\sum_{n}|\psi_{n}\rangle\langle\nabla\phi^{n}|,
\end{equation}
where the second equality is ensured by the completeness relation (\ref{right-left}). We then have
\begin{eqnarray}  \label{F-re} \nonumber
|\nabla \psi_j\rangle&=&{\text i}\mathbf{F}|\psi_j\rangle \\ \nonumber
\langle\nabla \phi^j|&=&-{\text i}\langle \phi^j|\mathbf{F}  \\  \nonumber
\nabla\times\mathbf{F}&=&-{\text i}\sum_{n}|\nabla \psi_{n}\rangle\times\langle \nabla \phi^{n}| \\ \nonumber
&=&-{\text i}\sum_{n}\mathbf{F}|\psi_{n}\rangle\times\langle\phi^{n}|\mathbf{F} \\
&=&-{\text i}\mathbf{F}\times \mathbf{F}.
\end{eqnarray}
In terms of $\mathbf{F}$ the Berry curvature can be expressed  as
\begin{equation}
\mathbf{B}_j={\text i} \sum_{j'} \langle \phi^j|\mathbf{F}|\psi_{j'}\rangle\times\langle \phi^{j'}|\mathbf{F}|\psi_j\rangle
={\text i}\langle \phi_j|\mathbf{F}\times \mathbf{F}|\psi_j\rangle.
\end{equation}
Finally, by virtue of $\nabla\times\mathbf{F}$ in Eq.~(\ref{F-re}), we find
\begin{eqnarray} \label{e41}  \nonumber
&&\nabla\cdot\mathbf{B}_j\\ \nonumber
&=&{\text i}[\langle\nabla \phi^j|\cdot(\mathbf{F}\times \mathbf{F})|\psi_j\rangle+
\langle \phi^j|(\mathbf{F}\times \mathbf{F})\cdot|\nabla\psi_j\rangle \\ \nonumber
&&+\langle \phi^j|\nabla\cdot(\mathbf{F}\times\mathbf{F})|\psi_j\rangle] \\ \nonumber
&=&{\text i}[-{\text i}\langle \phi^j|\mathbf{F}\cdot(\mathbf{F}\times \mathbf{F})|\psi_j\rangle
+{\text i}\langle \phi^j|(\mathbf{F}\times\mathbf{F})\cdot \mathbf{F}|\psi_j\rangle \\ \nonumber
&&+\langle \phi^j|(\nabla\times\mathbf{F})\cdot \mathbf{F}|\psi_j\rangle-\langle \phi^j|\mathbf{F} \cdot(\nabla\times\mathbf{F})|\psi_j\rangle] \\
&=&0.
\end{eqnarray}
This means that the divergence of the Berry curvature is always zero as long as the eigenstates are well-defined and smooth.
We define monopoles as the points in the parameter space,  where
the divergence of the Berry curvature becomes non-zero.   In the Hermitian systems,
the divergence of the Berry curvature is non-zero only at degenerate points, implying that in Hermitian systems
degenerate points and monopoles are equivalent.  The situation is very different for non-Hermitian systems,
where besides exceptional points (the equivalence of degenerate points) the Berry curvature
becomes discontinuous at branch cuts. As a result, the monopoles of non-Hermitian systems include
both exceptional points and branch cuts.

Let us now focus on the simple case of Hamiltonian (\ref{gene-H}) and use it to illustrate the above point.
At points away from the exceptional points, there are two well-defined eigenstates.
To compute Berry curvature over the whole parameter space, we need to choose one of them and at the same time
keep it change smoothly over the parameter space as far as possible. However, mathematically,
this smoothness can not be achieved in the entire parameter space and has to be disrupted at a branch cut.
As a result, the eigenstate becomes discontinuous at the branch cut and
the divergence of Berry curvature becomes non-zero.

\begin{figure}[t]
\includegraphics[width=1.0\linewidth]{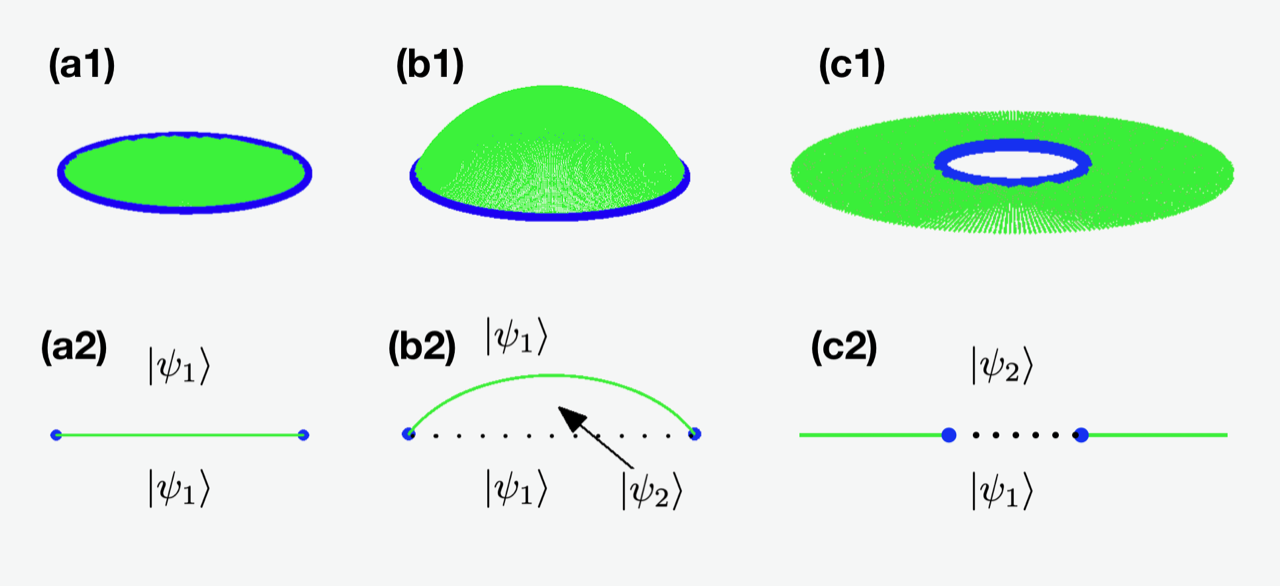}
\caption{(color online) Branch cuts (monopoles) for the non-Hermitian Hamiltonian (\ref{gene-H}).
(a1,b1,c1) are three different branch cuts. The blue circle consists of all the exceptional points.
The cuts of (a1,b1,c1) by the $(p_x,p_z)$ plane are shown in (a2,b2,c2), respectively,
where we show how the eigenstates are chosen in different regions for one branch.  }
\label{poles}
\end{figure}

As is well known, the choice of branch cut for a complex function is rather arbitrary as long as it originates from
a point where the complex function is ill-defined. For the complex function $\sqrt{\mathbf{p}^2-s^2+2\text{i}p_zs}$,
the branch cut can be any surface whose edge is the circle of exceptional points. It includes surfaces that extend
to infinity.

Three different choices of branch cuts are illustrated in  Fig. \ref{poles}.
In the first row (a1,a2), the branch cut is the given by  $p_x^2+p_y^2<s^2,p_z=0$. For one of the two branches,
we can choose eigenstates as $\ket{\psi_1}$. Note that $\ket{\psi_1}$ has different values
when the parameter approaches the branch cut from the above or the below.

In Fig.\ref{poles} (b1,b2), a different branch cut is shown. We denote the region enclosed by
the green dome and the disk ($p_x^2+p_y^2<s^2,p_z=0$)  as $\Omega$.
In this case, we can choose $\ket{\psi_2}$ in $\Omega$ and $\ket{\psi_1}$ elsewhere.
Note that $\ket{\psi_1}$ and $\ket{\psi_2}$ are smoothly connected at the dashed line in  Fig.\ref{poles} (b2).

In Fig.\ref{poles} (c1,c2),  the branch cut (monopole) is chosen to be $p_x^2+p_y^2>s^2,p_z=0$, i.e.,
the whole infinite $p_x-p_y$ plane outside the circle of exceptional points.
For this branch cut, we choose eigenstate $\ket{\psi_1}$ for $p_z<0$ and
eigenstate $\ket{\psi_2}$ for $p_z>0$.

It is worthwhile to note that the branch cut is totally irrelevant of the $GL(l,\mathbb{C})$ gauge transformation as shown in Eq.~(\ref{gauge55}). To define the monopole, we must choose one of the two non-Hermitian eigenstates given by (\ref{wf1}) and (\ref{wf2}) on each point in parameter space. The branch cut is necessary for the selection of eigenstate for the M\"obius distribution of eigenstate in parameter space. On the contrary, the $GL(l,\mathbb{C})$ gauge transformation is associated with the local self-contained simultaneous transformations of biorthonormal eigenstate and Berry connection, irrelevant of the selection of eigenstate and branch cut. As a result, the monopole composed of the exceptional points and branch cut is not affected by the $GL(l,\mathbb{C})$ gauge transformation.

We consider the case (a) in Fig.~\ref{poles}, where the branch cut is a disk given by  $p_x^2+p_y^2<s^2,p_z=0$.
It is clear that there is a discontinuity of the function $\sqrt{\mathbf{p}^2-s^2-2\text{i}p_zs}$ between $p_z>0$
and $p_z<0$ as $\mathbf{p}^2<s^2$, so the wavefunctions Eqs.~(\ref{wf1}) and (\ref{wf2}) and Berry curvatures~(\ref{Berry-Cur})
are discontinuous on the disk $p_x^2+p_y^2<s^2,~p_z=0$. We call this disk-like monopole
the natural monopole corresponding to the natural separatrix given in Eq.~(\ref{Gen}) (see Fig.~\ref{poles}(a)).

From Eq.~(\ref{Berry-Cur}), it can be found that the magnetic charge is distributed on the disk according to the density,
\begin{equation} \label{denn}
\rho_{1,2}=\pm \frac{s}{(s^2-p^2)^{3/2}},  \quad \text{for}\; p<s,
\end{equation}
where $p$ is the distance to the center of the disk (the origin), $+/-$ is for the first  /second  eigenstate.
When $p=s$, Eq.~(\ref{denn}) cannot apply. Consequently,  in order to derive the Chern number, one
has to integrate the Berry curvature over a closed surface enclosing the disk-like monopole. The resulted Chern number is
\begin{equation}
\mathcal{C}=\mp2\pi,
\end{equation}
where the sign $-/+$ is for the first ($+$) /second ($-$) eigenstate. It is worth noting that the Chern number cannot be obtained by integrating the density given by (\ref{denn}), because Eq.~(\ref{denn}) cannot apply as $p=s$ (the edge of the disk).

For a finite monopole, the Chern number can be easily calculated by integrating the Berry curvature over a closed surface
enclosing the monopole. For all the finite monopole the Chern's number is the same as that of the natural
one given in Fig.~3(a), i.e., $\mathcal{C}=\mp2\pi$. The reason for the same Chern
number lies in that a finite branch cut can never influence the Berry curvature in the infinite (outside the region $\Omega$).

Alternatively, one can make an infinite branch cut to label different eigenstates. There are infinite possibilities as for the finite branch cut.
The simplest case is  the plane infinite monopole given by $p_x^2+p_y^2\geq s^2,~p_z=0$ as plotted in Fig.\ref{poles} (c1,c2).
For such an infinite branch cut, the calculation of Chern's number becomes very different. We use the infinite branch cut
$p_x^2+p_y^2>s^2,~p_z=0$ as plotted in Fig.\ref{poles} (c1,c2).  The density of charge over the infinite plane can be calculated as
\begin{equation} \label{denn2}
\rho=\pm \frac{\text{i}s}{(p^2-s^2)^{3/2}},  \quad \text{for}\; p>s,
\end{equation}
where $p$ is the distance to the origin and  $+/-$ are for different choice of eigenstates.
 As Eq.~(\ref{denn2})  does not apply at $p=s$, we can not find the Chern number by integrating this charge density over the monopole.
The Chern's number of this infinite monopole can be derived by first considering the finite monopole $\Sigma(r)$ composed of,
\begin{eqnarray} \nonumber \label{Sigma}
\Sigma(r)&=&\{s^2\leq p_x^2+p_y^2\leq r^2,~p_z=0\}\\
&+&\{p_x^2+p_y^2+p_z^2=r^2,~~p_z\geq0\}.
\end{eqnarray}
According to the result that the Chern's number of a finite monopole is constantly $\mp2\pi$ ($\mp$ for the two different choice of eigenstate, respectively), we have
\begin{equation}
\mathcal{C}(\Sigma(r))=\mp2\pi.
\end{equation}
It is then necessary to work out the charge  of the semi-sphere,
\begin{equation}
S(r)=p_x^2+p_y^2+p_z^2=r^2,~~p_z\geq0\,.
\end{equation}
It can be done by integrating the corresponding Berry curvature given by (\ref{Berry-Cur}) over $S(r)$.
As the Berry curvature is discontinuous on $S(r)$, the charge is equal to the flux difference between in and out of the surface.
For one of the branch cuts, we have
\begin{eqnarray} \nonumber
\mathcal{C}(S(r))&=&\int_{S(r)} \left[d\mathbf{S}\cdot\mathbf{B}_1+d\mathbf{S}'\cdot\mathbf{B}_2 \right]\\
&=&\int_{S(r)} d\mathbf{S}\cdot\left[\mathbf{B}_1-\mathbf{B}_2 \right],
\end{eqnarray}
where $d\mathbf{S}$ is the directed surface element on $S(r)$ with the positive $p_z$ axis as the positive direction
while $d\mathbf{S}'=-d\mathbf{S}$ is the directed surface element with the negative $p_z$ axis as the positive direction.
The charge in the ring-belt-like region $s^2\leq p_x^2+p_y^2\leq r^2,~p_z=0$ is then
\begin{equation}
\mathcal{C}(\Sigma(r))-\mathcal{C}(S(r))=-2\pi-\mathcal{C}(S(r)).
\end{equation}
As $r\rightarrow\infty$, the total charge in the  semi-sphere tends to $\lim_{r\rightarrow\infty}\mathcal{C}(S(r))=-2\pi$ and the Chern number of the infinite monopole $p_x^2+p_y^2\geq s^2,~p_z=0$ is thus
\begin{equation}
\lim_{r\rightarrow\infty} [\mathcal{C}(\Sigma(r))-\mathcal{C}(S(r))]=-2\pi-(-2\pi)=0.
\end{equation}

For the alternative choice of eigenstate (choose eigenstate $\ket{\psi_2}$ for $p_z<0$ and
eigenstate $\ket{\psi_1}$ for $p_z>0$.), we have
\begin{eqnarray}
\mathcal{C}(S(r))=\int_{S(r)} d\mathbf{S}\cdot\left[\mathbf{B}_2-\mathbf{B}_1 \right]\,.
\end{eqnarray}
 The Chern number  in this case is
\begin{equation}
\lim_{r\rightarrow\infty} [\mathcal{C}(\Sigma(r))-\mathcal{C}(S(r))]=2\pi-2\pi=0.
\end{equation}

The Chern number (together with the Berry curvature) is apparently irrelevant of the $GL(l,\mathbb{C})$ gauge transformation, since the monopole itself is only determined by the branch cut.


\section{Summary}
To summarize, we have studied the  monopole of non-Hermitian quantum systems and
found that in non-Hermitian systems the monopoles
contain not only exceptional points but also branch cuts.
The monopole in non-Hermitian quantum mechanics  thus depends on the choice of branch cuts,
which is similar  to the choice of gauges.
Our results indicate that the Berry curvature is a more fundamental geometric quantity than the monopole and Chern number.
Although we have so far only considered two-mode non-Hermitian systems, the results should be generic.
For non-Hermitian systems with more than two modes, we expect that branch cuts
become more complicated and  the M\"obius loop
be replaced by loops where one can not get back to the original eigenstates even after traversing it twice.
Since such non-Hermitian dynamics can be generically found or constructed in various physical systems, we expect
that our study offer new insights into the dynamical and topological properties of non-Hermitian systems.

\begin{acknowledgments}
 This work was supported by the The National Key R\&D Program
of China (Grants No.~2017YFA0303302, No.~2018YFA0305602) .
\end{acknowledgments}

\end{document}